\documentclass[twocolumn,10pt]{IEEEtran}

\usepackage{cite}
\usepackage{graphicx}
\usepackage{amsmath}
\usepackage{amsfonts}
\usepackage{epstopdf}
\usepackage{xcolor}
\usepackage{amssymb,bm,upgreek}
\usepackage{algorithm}
\usepackage{algorithmic}
\usepackage{multirow}
\usepackage{transparent}
\usepackage{subfigure}
\usepackage{import}

\DeclareMathOperator\erf{erf}

\newcommand{\Prob}{\textnormal{Pr}}
\newcommand{\m}{\textnormal{m}}
\newcommand{\s}{\textnormal{s}}
\newcommand{\ob}{\textnormal{ob}}
\newcommand{\bit}{\textnormal{bits}}
\newcommand{\inter}{\textnormal{slot}}
\newcommand{\Pe}{\textnormal{Pe}}

\newtheorem{remark}{Remark}

\newtheorem{corollary}{Corollary}

\begin{document}

\title{Molecular Information Delivery in Porous Media}

\author{Yuting Fang,
        Weisi Guo,
        Matteo Icardi,
        Adam Noel,
        and Nan Yang\vspace{-8mm}

\thanks{Y. Fang and N. Yang are with the Research School of Electrical, Energy and Materials Engineering, The Australian National University, Canberra, ACT 2600, Australia (e-mail: \{yuting.fang, nan.yang\}@anu.edu.au).}
\thanks{W. Guo and A. Noel are with the School of Engineering, University of Warwick, Coventry, CV4 7AL, UK (e-mail: \{weisi.guo, adam.noel\}@warwick.ac.uk).}
\thanks{Matteo Icardi is with the School of Mathematical Sciences, University of Nottingham, Nottingham, NG7 2RD, UK (e-mail:Matteo.Icardi@nottingham.ac.uk)}
\thanks{This work is partly funded by US AFOSR grant FA9550-17-1-0056.}
}

\maketitle

\begin{abstract}
Information delivery via molecular signals is abundant in nature and potentially useful for industry sensing. Many propagation channels (e.g., tissue membranes and catalyst beds) contain porous medium materials and the impact this has on communication performance is not well understood. {Here, communication through realistic porous channels is investigated for the first time via statistical breakthrough curves. Assuming that the number of arrived molecules can be approximated as a Gaussian random variable and using fully resolved computational fluid dynamics results for the breakthrough curves, the numerical results for the throughput, mutual information, error probability, and information diversity gain are presented. Using these numerical results, the unique characteristics of the porous medium channel are revealed.}
%Here, communication through realistic porous channels is analyzed for the first time via statistical breakthrough curves. Assuming that the number of arrived molecules can be approximated as a Gaussian random variable, analytical results for the throughput, mutual information, error probability, and information diversity gain are presented. Using numerical results, the unique characteristics of the porous medium channel are investigated
%The potential for future research about exploring the communication performance via porous medium is also discussed. and their potential for diversity gain is demonstrated for different P\'{e}clet number fluid dynamic regimes.
\end{abstract}

%\begin{IEEEkeywords}
%\end{IEEEkeywords}

\IEEEpeerreviewmaketitle
\vspace{-6mm}
\section{Introduction}\label{sec:intro}

For decades, conveying information over a distance has been an important component of organized behavior. The conventional electromagnetic signals are not appropriate in many biological and chemical engineering environments since electromagnetic signals quickly decay in such environments.  %liquid/obstacle fields and the unwanted effects caused by the electromagnetic signals may pose a health risk.
In nature, molecular signals are used for many microorganisms to signal each other and share information, e.g., quorum sensing and excitation-contraction coupling \cite{Nakano2005}. Inspired by nature, molecular communication (MC) has been proposed. %where a transmitter (TX) releases small particles into an environment and the RX detects the information from the molecules arrived.
%where the molecules propagate until they arrive at a receiver (RX),

Significant research has been done to investigate molecular signal propagation in both free space (FS) and simple bounded environments, e.g., \cite{Nakano2012, Guo15Obstacle}. These papers have been suitable for establishing tractable limits on communication performance by assuming %non-obstacle propagation environment.
that molecules propagate in environments without obstacles.
However, in many biological (e.g., tissue membrane \cite{Rittmann1993}) and chemical engineering (e.g., catalyst bed \cite{Perego2012}) environments, the channel consists of porous medium (PM) materials. The PM is a solid with pores (i.e., voids) distributed more or less uniformly throughout the bulk of the body \cite{Bear2013}. Many natural and man made substances, e.g., rocks, soils, and ceramics, can also be classified as PM materials \cite{Scheidegger1961}.

PM channels are fundamentally different from FS channels due to the intricate network of pores. The molecules undergo complex trajectories and experience heterogeneous advection as they propagate through pores of different sizes and lengths, causing so-called mechanical dispersion \cite{Bear2013,Icardi:2014aa}, which is an augmented effective diffusion caused by velocity fluctuations. More importantly, particles may become trapped in immobile or re-circulation zones in the vicinity or the wake of solid grains \cite{Eleonora2016,dentz_icardi_hidalgo_2018}, therefore taking some time to exit, and causing non-trivial anomalous transport phenomena, such as long tails in the arrival time distributions. Hence, it is of fundamental importance to investigate what impact these PM flow and transport properties have on the MC performance.
%porous channels to increase the path diversity due to its complex pore structure and hence both improve the diversity gain while also increasing the impact of . %Recently, \cite{Murin2018} studied diversity in one-shot MC over molecular timing channels where

%In this paper, we consider MC via the PM channel.
This work is the first to consider a PM channel in MC. We consider a binary sequence transmitted between a transmitter (TX) and a receiver (RX) located at the ends of the PM channel. The main contributions are summarized as follows:
\begin{enumerate}
\item {Assuming that the number of molecules arrived can be approximated as a Gaussian random variable (RV), we present numerical results for different performance metrics, i.e., throughput, mutual information, and error probability, for the channel using fully resolved computational fluid dynamics results for the breakthrough curves.} We also numerically evaluate the diversity gain that is defined (as in \cite{Murin2018}) as the exponential decrease rate of the probability of error as the number of released molecules increases.

\item Using numerical results, we investigate the differences in channel characteristics and performance metrics between a PM and diffusive FS channel with flow. {In particular, we show that the tail of the PM channel response is longer than that of the FS channel, which can significantly affect the communication performance, e.g., the inter-symbol interference (ISI) in the case of concentration-modulated transmission is more severe.\footnote{{The long tails in the arrival time distribution do not necessarily mean the existence of ISI. For example, when timing-based modulation is considered, the long tail of channel response leads to transposition errors \cite{7940049}.}}}
\end{enumerate}

The rest of this paper is organized as follows. The system model is presented in Section \ref{sec:system}. The performance metrics are derived in Section \ref{sec:metrics}. Numerical results are presented in Section \ref{Sec:results}. The paper is concluded in Section \ref{sec:con}.

\vspace{-4mm}
\section{System Model}\label{sec:system}

\begin{figure*}[!t]
\centering
\subfigure[]{\label{fig:system}\includegraphics[width=2.75in]{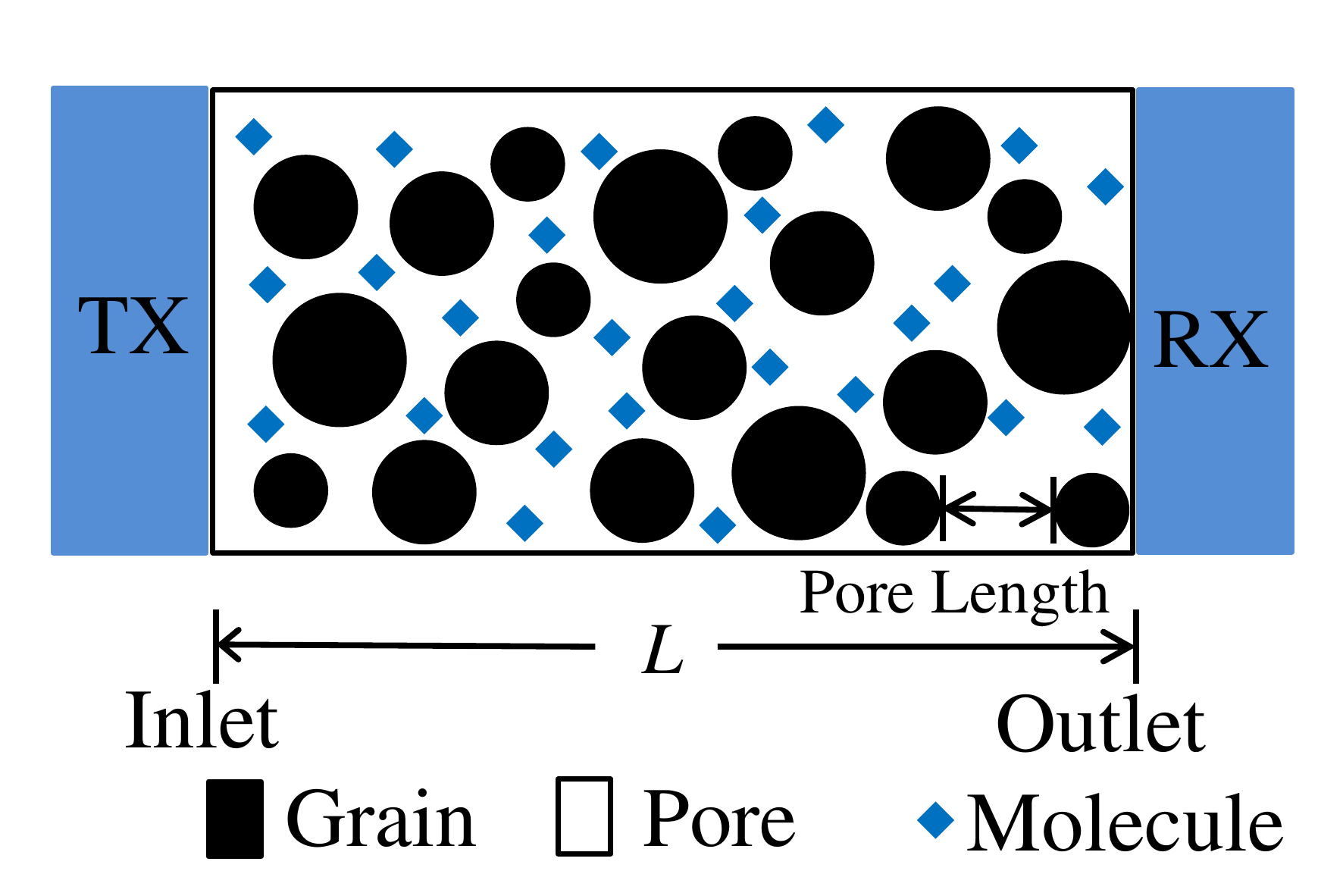}}
\subfigure[]{\label{fig:porous}\includegraphics[width=1.4in]{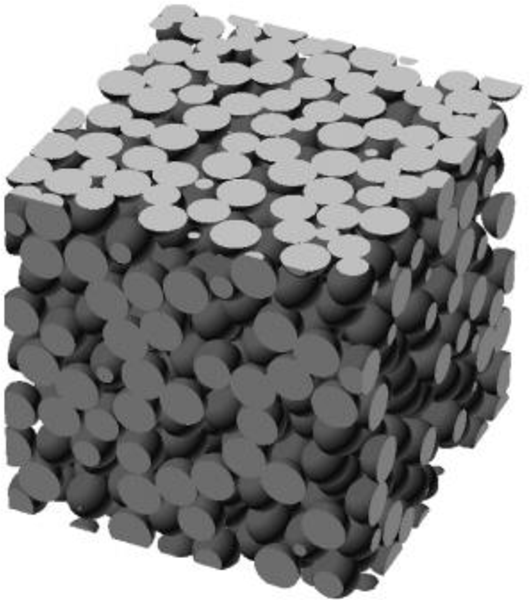}}
\subfigure[]{\label{fig:motion}\includegraphics[width=2in]{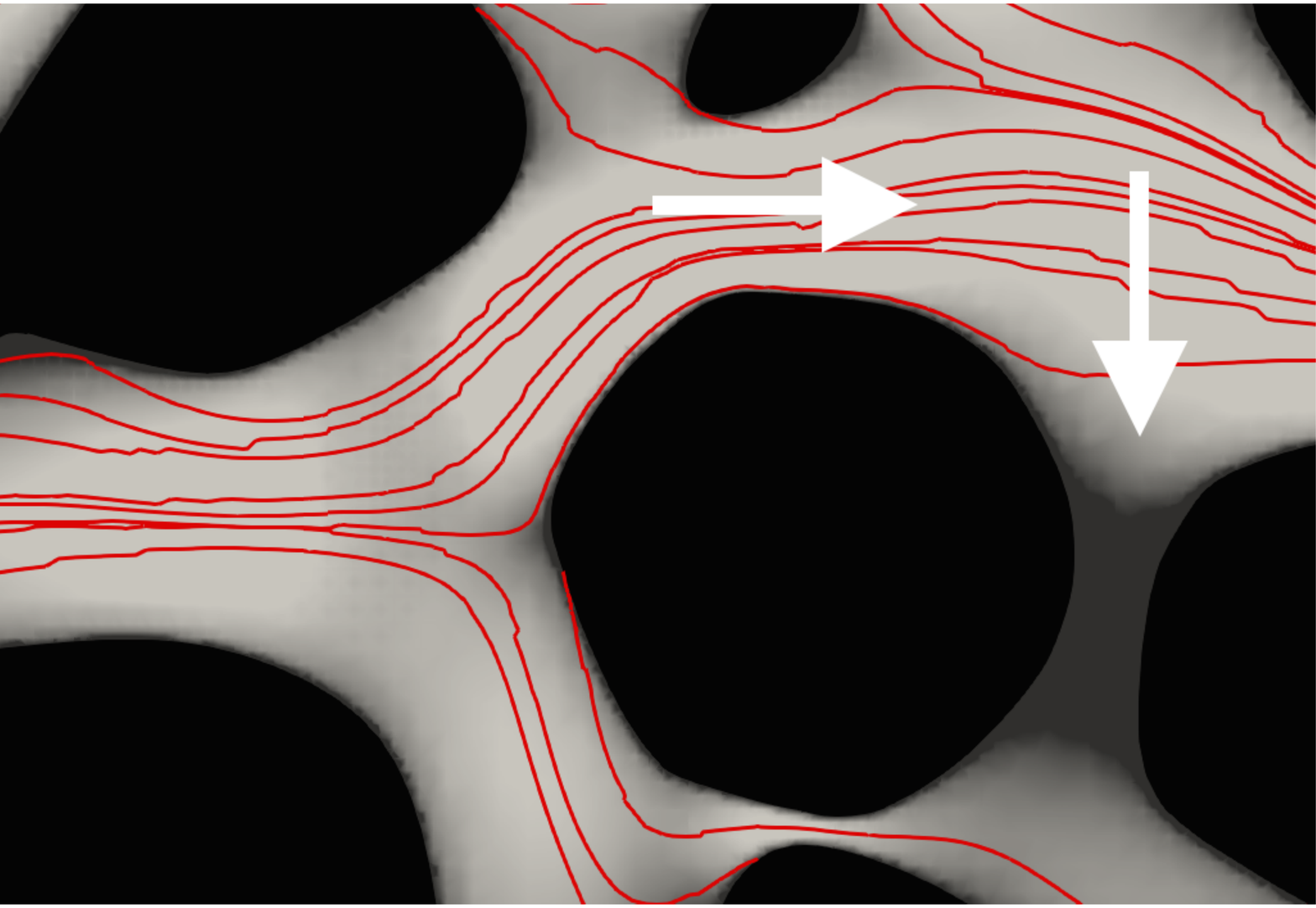}}\vspace{-2mm}
\caption{\subref{fig:system}: A 2D sketch of the considered system model, where $L$ is the distance between the TX and RX. \subref{fig:porous}: A 3D sample of a PM \cite{CHUEH2014183}. \subref{fig:motion}: Illustration of molecular transport through a PM with heterogeneous advection \cite{dentz_icardi_hidalgo_2018}, where the red lines represent streamlines of the laminar flow; the shading of the background denotes the flow velocity which decreases from light to dark; the horizontal arrow denotes transport of molecules over the length of a pore in streamwise direction; and the vertical arrow indicates transport of molecules across streamlines into low velocity zones in the wake of the solid grains. In \subref{fig:porous} and \subref{fig:motion}, the grains are represented in grey and black, respectively.}\label{figures}\vspace{-6mm}
\end{figure*}

%\begin{figure}[!t]
%\centering
%\includegraphics[width=1\columnwidth]{system_modelv3.pdf}
%\caption{THIS SHOULD BE CHANGED. THE ARROWS DESCRIBING $\vec{v}$ SHOULD NOT BE PARABOLIC. ON THE CONTRARY THEY SHOULD CLEARLY GIVE THE IDEA OF A COMPLICATED FLOW FIELD, THAT FOLLOWS THE SHAPE OF THE GRAINS AND, AT THE SAME TIME, HAVE A VERY WIDE RANGE OF VELOCITY MAGNITUDES. Two-dimensional sketch of the considered system model, where $L$ is the distance between the TX and the RX.}
%\label{fig:system}
%\vspace{-4mm}
%\end{figure}

%\begin{figure}[!t]
%\centering
%\includegraphics[width=0.9\columnwidth]{porous3D.png}
%\caption{Three-dimensional (one eighth) sample of a PM and a two-dimensional slice of it %\cite{dentz_icardi_hidalgo_2018}.}
%\label{fig:porous}
%\vspace{-4mm}
%\end{figure}

We consider an MC system via the PM in a three-dimensional (3D) environment where the TX and the RX are located at the inlet and the outlet of the PM, respectively. A two-dimensional (2D) sketch of the considered system is given in Fig. \ref{fig:system} and a 3D sample of a PM is shown in Fig. \ref{fig:porous}.
In a PM, pores and grains refer to its void and solid components, respectively. Grain size distribution and porosity (i.e., the ratio of the volume of voids over the total volume) affect the transport behavior in the PM. In the following, we detail the key steps of the considered system.

\emph{\underline{Modulation and Emission}}: A sequence of binary symbols is transmitted with $\Prob(X_n=1)=P_1$, where $X_n$ is the $n$th transmitted symbol. %since they are required to convey complex information, such as blood pressure, PH value, and target location, in the potential MC applications (e.g., abnormality detection and drug delivery) \cite{Farsad2016}.%e.g., health monitoring, targeted drug delivery, and lab-on-a-chip
We consider the on-off keying modulation scheme with a fixed symbol slot length $T$, which is commonly adopted in MC literature, i.e., at the beginning of the $n$th symbol slot, the TX releases $N$ molecules if $X_n=1$; otherwise, no molecule is released. The TX uniformly releases the molecules over the cross section at the inlet of the PM. {We note that the use of a binary sequence is expected in MC between nanomachines to exchange the amount of information required for executing complex collaborative tasks, e.g., disease detection \cite{DBLP:journals/corr/abs-1901-05546}, and binary symbols are easier to transmit than symbols that carry more bits of information.}

%Using the on-off keying modulation scheme, with probability $\Prob(X_n=1)=P_1$ and when $X_n=0$.

%\begin{figure}[!t]
%\centering
%\includegraphics[width=0.8\columnwidth]{porous_motion.pdf}
%\caption{ Illustration of molecular transport through the PM with heterogeneous advection. The thin red lines represent streamlines of the laminar flow in a cross-section of the PM. The shading of the background denotes the flow velocity which decreases from light to dark. Black areas denote the grains. The horizontal arrow denotes transport of molecules over the length of a pore in streamwise direction and the vertical arrow indicates transport of molecules across streamlines into low velocity zones in the wake of the solid grains \cite{dentz_icardi_hidalgo_2018}.}
%\label{fig:motion}
%\vspace{-4mm}
%\end{figure}

\emph{\underline{Transport through the PM}}: We consider the PM filled with an incompressible fluid of viscosity $\mu$, moving with a mean velocity $\vec{v}_m$ oriented from the TX to the RX.
Due to the small pore sizes, the flow is laminar (Reynolds number of the flow is negligible) and governed by the Stokes equation $\mu\nabla^2\vec{v}(\mathbf{a})  =\nabla p(\mathbf{a})$
%\begin{equation}\label{flow}
%\mu\nabla^2\vec{v}(\mathbf{a})  =\nabla p(\mathbf{a})\, ,
%\end{equation}
together with the incompressibility condition $\nabla\vec{v}(\mathbf{a})  =0$,
%\begin{equation}\label{flow1}
%\nabla\cdot\vec{v}(\mathbf{a})  =0\, ,
%\end{equation}
where $\nabla$ is the nabla operator, $\mathbf{a}$ denotes location, $\vec{v}$ is the velocity, and $p(\mathbf{a})$ is the pressure. The boundary conditions are of zero velocity (no-slip) on the surface of the solid grains, and periodic on the external boundaries, with a fixed pressure gradient along the mean flow direction. The resulting velocity field $\vec{v}$ is characterized by a chaotic heterogeneous structure. {Mechanical entrapment
of molecules may occur in PM when the molecules are too
large to enter small pores \cite{Sameer2018}. Small molecules such as water and salt molecules can travel through PM, but large molecules
such as polymer molecules will be trapped and accumulate
in these small pores. Although these effects are not explicitly
modelled here (they would, in fact, require a Lagrangian
description of molecules as rigid bodies), a similar effect
is here included when the flow velocity is high compared
to molecular diffusion. The few molecules that diffuse into
stagnant regions can get trapped for relatively long times
before diffusing back into the main flow channels.}

The molecular transport in the pores is due to molecular diffusion and the complex heterogeneous advection around solid grains, as shown in Fig. \ref{fig:motion}.
The molecular concentration $c(\mathbf{a},t)$ is modeled by an advection-diffusion equation \cite{Bejan2013}:
\begin{equation}\label{transport}
{\partial c(\mathbf{a},t)}/{\partial t}+\vec{v}(\mathbf{a})\nabla c(\mathbf{a},t)-D\nabla^2c(\mathbf{a},t)=0\, ,
\end{equation}
where $D$ is the constant diffusion coefficient, with a constant flux of molecules on the inlet and zero diffusive flux on all other boundaries. %is mediated by the laminar flow (i.e., heterogeneous advection) and Brownian motion (i.e., molecular diffusion). These transport mechanisms give rise to the hydrodynamic dispersion. Dispersion quantifies the impact of velocity fluctuations on molecule transport.
Although these equations are linear and relatively easy to solve, the complexity of the geometry makes the discretization and solution particularly cumbersome\cite{Icardi:2014aa}.

The P\'{e}clet number (Pe), which compares advective and diffusive transport over the whole PM length $L$, is given by $\Pe = {|\vec{v}_m|L}/{D}$.
%\begin{align}\label{peclet}
%Pe = \frac{|\vec{v}_m|L}{D}.
%\end{align}	
Thanks to the interplay of these two phenomena, molecules not only are transported along the streamlines but also travel across streamlines, experiencing therefore a wide range of velocities, and possibly reaching stagnant
%or low velocity
zones in the wake of the solid grains.
%Transport along streamlines are due to mean advection in the pore and diffusion. Transport across the streamlines are due to diffusion.
Molecules that enter these zones can remain there for some time before they escape and return into the mobile portion of the medium. {The transport of molecules through PM may also be affected by electro-chemical effects. For example, molecules may contain polar groups, which will attach to the available polar points on the PM surface \cite{Sameer2018}. Depending on the PM surface net ionic charge, electrostatic attraction or repulsion would occur for ionic molecules, which enhance or reduce the ionic molecular adsorption on the surface of PM. For the tractability of the distribution of first arrival time of molecules, we do not consider electrical effects on molecular propagation.}
%The importance of these effects to the overall transport properties have been recently identified in \cite{dentz_icardi_hidalgo_2018}.
%Therefore, the motion of molecules is mediated by interaction of the intra- and interpore molecule advection and diffusion mechanisms.

\emph{\underline{Reception and Demodulation}}: We consider a RX that is mounted on the cross section at the outlet of the PM and is able to count the number of molecules that arrive. %molecules are absorbed by the RX once they arrive the outlet.
To decrease the complexity, we consider a fixed threshold-based demodulation rule at the RX: $Y_n=1$ if $N^{\ob}_{n}\geq\xi$; otherwise, $Y_n=0$, where $Y_n$ is the $n$th received symbol, $N^{\ob}_{n}$ is the number of molecules that arrive during the $n$th slot, and $\xi$ is a fixed threshold. {The transmission and reception of multiple symbols is possible. The encoding function at the TX can be implemented by synthesizing logic gates \cite{Moon2012GeneticPC}. A metabolic pathway of a biological cell can be synthesized into the TX to release specific molecules \cite{Chen2005ArtificialCC}. The computational processing at the RXs can be implemented based on \cite{Pischel,Silva}. The time synchronization between the TX and the RXs can be implemented using various methods, e.g., a blind synchronization algorithm \cite{ShahMohammadian2013} and quorum sensing-based method \cite{Abadal2011}.}

%Since this work focuses on exploring the channel characteristics of a porous medium, for simplicity we do not consider an adaptive threshold \cite{Chang2018}. %which will be considered in the future work to improve the error performance of the system.
%\begin{align}\label{detectorRX}
%Y_n=
%\begin{cases}
%1,&\mbox{if $N^{\ob}_{n}\geq\xi$,}\\
%0,&\mbox{otherwise}.
%\end{cases}
%\end{align}
%We assume that at the outlet during each symbol slot, and chemically reacts to those molecules to demodulate information at the end of each symbol slot.
\vspace{-4mm}
\section{Performance Metrics}\label{sec:metrics}
In this section, we present the analytical results of system performance metrics. To this end, we first analyze the (cumulative) breakthrough curve, i.e., the cumulative density function (CDF) of the first arrival time at the outlet of any molecule released from the inlet, which is used for characterizing molecular transport in the PM. This is given by \cite{dentz_icardi_hidalgo_2018}
\begin{align}\label{breakthrough}
F(t) \!=\! \frac{\int\!\int\!c(a_1\!=\!L,a_2,a_3,t)|\vec{v}(a_1=L,a_2,a_3)|da_2da_3}{\int\int |\vec{v}(a_1=L,a_2,a_3)|da_2da_3},\vspace{-4mm}
\end{align}	
where $\mathbf{a}=\{a_1,a_2,a_3\}$ denotes location in Cartesian coordinates. The analytical expression for $F(t)$ is mathematically intractable, so we will rely on a numerical solution obtained by the full discretization of \eqref{transport} and \eqref{breakthrough}. For more details about numerical solvers, we refer the readers to \cite{Icardi:2014aa}.

{If the TX and RX only partially cover the media inlet and outlet, then the breakthrough curve needs to be re-computed since the boundary conditions change and the dimension of the problem effectively increases (since the whole coverage case is effectively a one-dimensional system). More generally, when the TX and the RX are located arbitrarily in an open three-dimensional domain, one would need to consider a full non-diagonal and anisotropic dispersion tensor \cite{Bear2013} and not only the longitudinal dispersion studied here. We expect that the difference between breakthrough curves with full and partial coverage, to resembles the difference between diffusion processes in one and more dimensions.}
%the mutual information between channel input and output (i.e., PM inlet and outlet), throughput, error probability, and diversity gain

\begin{remark}\label{perfo}
%Since we consider conventional modulation, emission, reception, and demodulation MC schemes for the PM channel, the derivation of performance metrics is relatively straightforward.
Assuming that the number of molecules arrived can be approximated as a Gaussian RV, we derive the mutual information $I$, throughput $C$, and error probability $Q$. Using particle-based simulation methods, \cite{NOEL201744,Yilmaz2015} have verified the accuracy of Gaussian approximation. According to the central limit theorem, the accuracy of this approximation improves as $N$ increases. Due to the space limitation, we present the derivation of statistical distributions of molecules arrived and system performance metrics in Appendix A.
\end{remark}

We next discuss the diversity gain. Each molecule behaves independently and experiences different propagation paths. Thus, the channel can be seen as a multiple-input and multiple-output channel and the RX achieves diversity when $N$ molecules are released. Also, there is an optimal $\xi$ that minimizes error probability $Q$, i.e., $Q^{\ast}=\underset{\xi}{\text{min}}\;Q$, where $Q^{\ast}$ is the optimal error probability. We define the \emph{diversity gain} as the exponentially decreasing rate of $Q^{\ast}$ as a function of increasing $N$. That is to say, if we can well approximate $Q^{\ast}$ with a form of $Q^{\ast}\approx\exp(-\alpha N+\beta)$, then $\alpha$ is the diversity gain. %The diversity gain quantifies the decreasing rate of $Q^{\ast}$ with $N$, i.e., a higher $\alpha$ means a higher decreasing rate and $\alpha=0$ means $Q^{\ast}$ does not change with $N$.
{The assessment of the diversity gain of different channels indicates which channel is more sensitive to the increase in the number of molecules released, without the need for explicitly calculating the probability of error. Specifically, if a higher diversity gain is achieved, the channel is more sensitive. Thus, the evaluation of diversity gain provides information regarding the fundamental properties of different channels, which for example facilities the appropriate selection of the number of molecules released for MC system design.} Since an explicit expression for $\alpha$ is mathematically intractable, we use a data-fitting method to obtain $\alpha$. The method will be detailed in Sec. \ref{Sec:results}. We note that a similar definition of $\alpha$ was studied in \cite{Murin2018} for timing channels, but our method for evaluating $\alpha$ is different from \cite{Murin2018}.

For $P_1=\frac{1}{2}$, we have following corollaries on $Q$ and $I$:
\begin{corollary}\label{limit}
The optimal error probability converges to zero when the released number of molecules for symbol ``1'' tends to infinity, i.e., $\lim_{N\to\infty} Q^{\ast} =0$.
\end{corollary}
%\begin{IEEEproof}
%The proof of Corollary \ref{limit} is given in Appendix \ref{app1}.
%\end{IEEEproof}
\begin{corollary}\label{mutual}
The mutual information is bounded by $I\leq1\,{\bit}/{\inter}$ and $I=1\,{\bit}/{\inter}$ is obtained if and only if $Q\to0$.
\end{corollary}
\begin{IEEEproof}
The proofs of Corollaries \ref{limit} and \ref{mutual} are given in Appendices B and C, respectively.
\end{IEEEproof}
\vspace{-4mm}
\section{Numerical Results}\label{Sec:results}
In this section, we present numerical results to investigate the channel response and communication performance of MC via the PM. {We consider the 3D sand-like PM described in \cite{Icardi:2014aa,dentz_icardi_hidalgo_2018}. The medium was generated according to the characteristics of standard sand samples. Specifically, the PM is a cube of size $L=2{\m}\m$, which is of the size of a representative elementary volume in terms of the definition of volumetric porosity \cite{Bear2013}. The typical porosity of many kinds of soils, e.g., gravel, sand, silt, and clay, is between $20\%$ and $50\%$, based on \cite{Das2008,SN670,Hough}. The typical grain diameter for medium sand is between $0.25\,{\m}\m$ and $0.5\,{\m}\m$ \cite{doi:10.1111/j.1365-3091.1966.tb01572.x}. We consider the porosity of $35\%$ and the grain diameter of $0.277\,{\m}\m$, which are within the normal range for sand samples.} The grain size distribution follows a Weibull distribution with Weibull parameter $k=7$. We also consider $n=10$ symbols are transmitted with $P_1=\frac{1}{2}$. The other parameters are given in Table \ref{tab:table1}. With these parameters, \cite{dentz_icardi_hidalgo_2018} numerically solved \eqref{transport} and \eqref{breakthrough}, obtaining the values of $F(t)$ for $\Pe = 3,30,300,1000$. The results in the following figures are obtained based on this simulation data. {In Fig~\ref{fig:mutualError}, Fig~\ref{fig:error_numMols}, and Table~\ref{tab:table2}, we assume that $N^{\ob}_{n}$ is a Poisson RV since we consider $N\leq100$. In Fig~\ref{fig:throughput}, we assume $N^{\ob}_{n}$ is a Gaussian RV since we consider $N=10^5$.}
%with length $L=2\,{\m}\m$ and contains approximately $2\times10^3$ grains. The average grain diameter is $\overline{d} = 0.277\,{\m}\m$. We also consider that the characteristic pore length $\ell_0$ is on the order of the average grain size, i.e., $\ell_0\approx\overline{d}=0.277\,{\m}\m$, which leads to
\begin{table}[!t]
\renewcommand{\arraystretch}{1.1}
\centering
\caption{Environmental Parameters}\label{tab:table1}\vspace{-2mm}
\begin{tabular}{c||c|c}
\hline
\bfseries Parameter &  \bfseries Symbol&  \bfseries Value \\
\hline\hline
Length of PM & $L$ & $2\,{\m}\m$ \\\hline
Number of grains & $\phi$ & $2\times10^3$\\\hline
Average grain diameter & $\overline{d}$ & $0.277\,{\m}\m$\\\hline
Characteristic pore length (estimated) & $\ell_0$ & $0.277\,{\m}\m$\\\hline
Mean velocity & $|\vec{v}_m|$ &$5.73\times10^{-6} \m/\s$\\
%Number of symbols transmitted & $n$ & $10$\\
\hline
\end{tabular}
\vspace{-3mm}
\end{table}

In order to provide more insights, we compare with a one-dimensional (1D) diffusive FS channel with a flow oriented from the TX to the RX, which is referred to as the ``FS channel'' in the following for brevity. This is because the PM channel is effectively a 1D channel due to the TX and RX covering the entire inlet and outlet.
The probability density function (PDF) of the first arrival time at $a_1=L$ in the FS channel is given by $f(t)= \frac{L}{\sqrt{4\pi Dt^3}}\exp(-\frac{(|\vec{v}_m|t-L)^2}{4Dt})$ \cite{Crank1980}.
%\begin{align}\label{1DPDF}
%f(t)= \frac{L}{\sqrt{4\pi Dt^3}}\exp(-\frac{(|\vec{v}_m|t-L)^2}{4Dt}).
%\end{align}
For this FS channel, we consider the same parameter values as those for the PM channel for the fairness of our comparison.
%length $L$, the diffusion coefficient $D$, mean velocity $|\vec{v}_m|$
\vspace{-6mm}
\subsection{Channel Response}
%In this subsection, we investigate the CDF and the PDF of the arrival time of the molecules in the PM and FS channel models.
\begin{figure}[!t]
\centering
\includegraphics[height=2in]{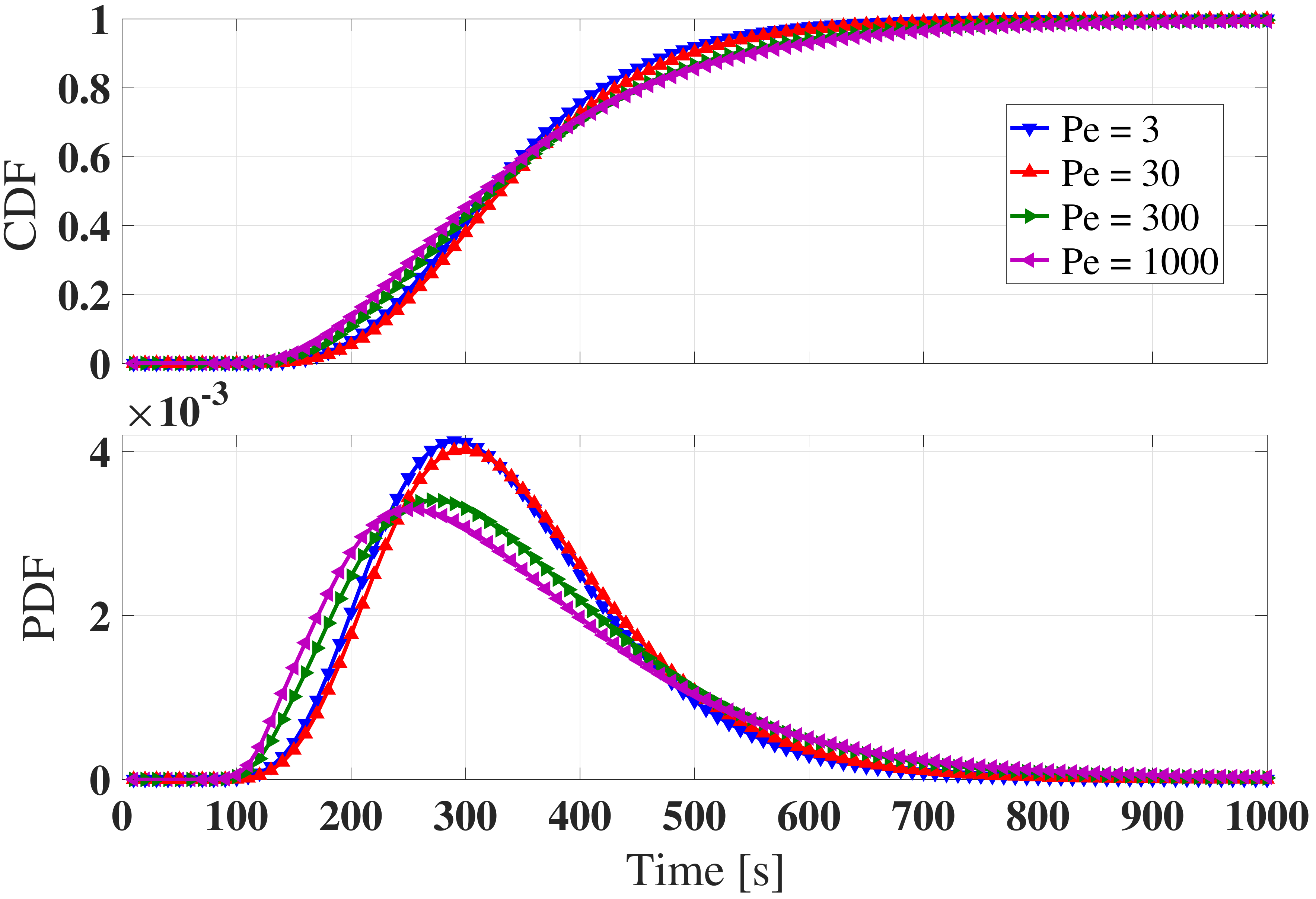}\vspace{-2mm}
\caption{The CDF  and PDF $f(t)$ of the arrival time of the molecule versus time $t$ in the PM channel for different $\Pe$. }
\label{fig:PDF}
\vspace{-4mm}
\end{figure}

%\begin{figure}[!t]
%\centering
%\includegraphics[height=2.4in]{PDF_Time}
%\caption{The PDF of the arrival time of the molecule $\frac{\partial F(t)}{\partial t}$ versus time $t$ in the PM channel for different $\Pe$: $\Pe = 3,30,300,1000$.}
%\label{fig:PDF}
%%\vspace{-4mm}
%\end{figure}

In Fig. \ref{fig:PDF}, we show the arrival time distribution in the PM channel. The PDF curves are obtained by numerically evaluating the derivative of $F(t)$. Firstly, for all $\Pe$, $F(t)\to1$ as $t\to\infty$, which means that all molecules released will eventually arrive at the RX. This is because no flow is going out of the lateral directions and no molecule can escape from the lateral directions by advection nor by diffusion. Secondly, when $\Pe$ is smaller, the CDF converges more quickly to 1, meaning that less molecules stay trapped in the PM.
%We observe that when $\Pe$ is smaller, the PDF curve has a higher peak value and shorter tails.the propagation environment is bounded due to immobile zones of the PM

\begin{figure}[!t]
\centering
\includegraphics[height=2in]{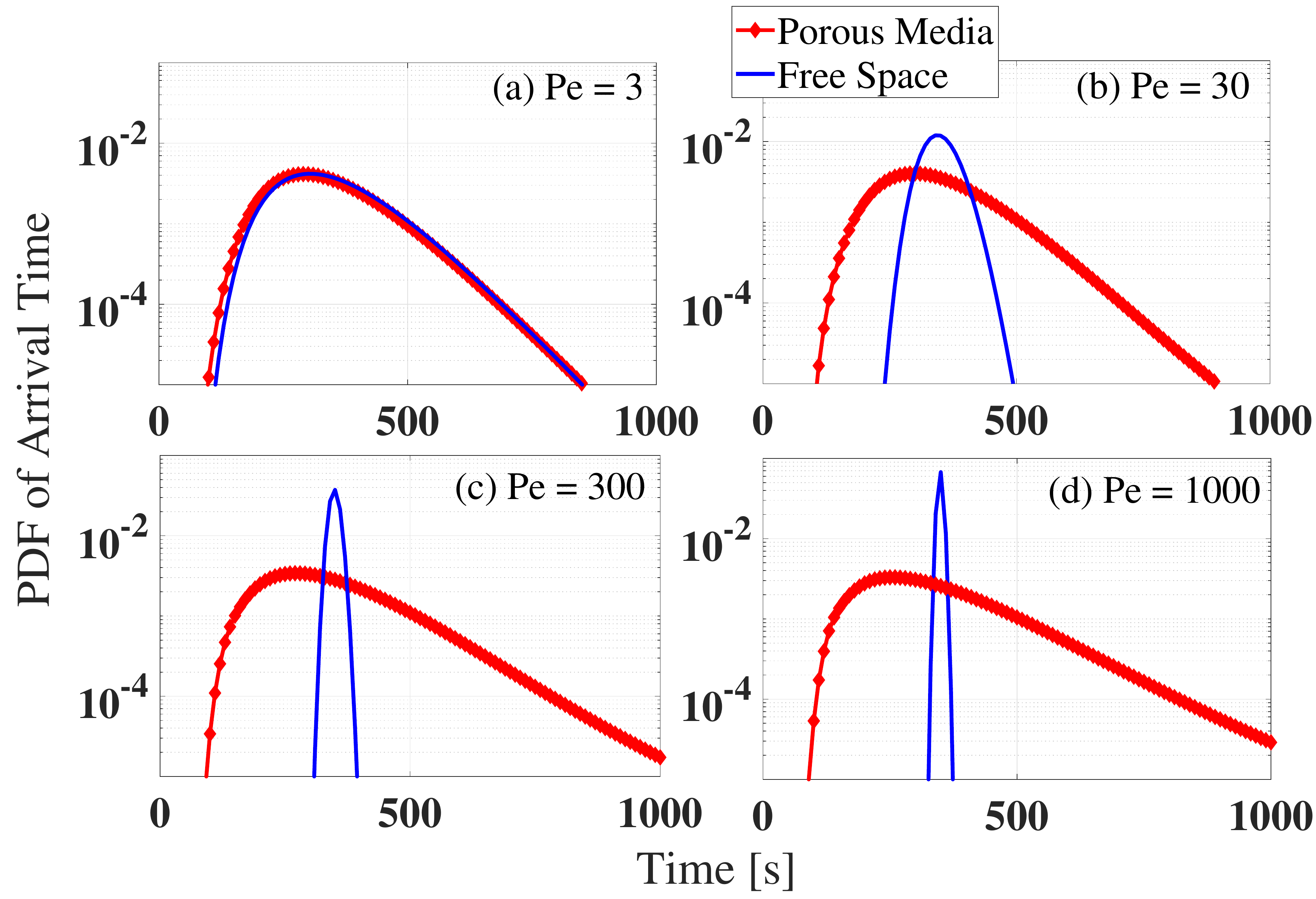}\vspace{-2mm}
\caption{The PDF $f(t)$ of the arrival time of the molecule versus time in the PM and FS channels for different $\Pe$.}
\label{fig:PDFCompare}
\vspace{-4mm}
\end{figure}

%\begin{figure}[!t]
%\centering
%\includegraphics[height=2.4in]{CDF_Time_1D_po}
%\caption{The CDF of the arrival time of the molecule versus time in the PM channel and FS channel for different $\Pe$: (a) $\Pe = 3$, (b) $\Pe = 30$, (c) $\Pe = 300$, and (d) $\Pe = 1000$.}
%\label{fig:CDFCompare}
%%\vspace{-4mm}
%\end{figure}

In Fig. \ref{fig:PDFCompare}, we compare the arrival time PDF in the PM channel with that in the FS channel. Interestingly, when $\Pe$ is 3, the PDF curve for the PM is similar to that for FS. This is because the fact that molecular diffusion is fairly large, causing particles to uniformly sample the velocity space, and resulting in an overall transport that can be conveniently described as a single advection-diffusion channel. Secondly, PM channel behavior is much less sensitive to $\Pe$ than in the FS channel. This is due to molecules entering dead-end pores or stagnant regions, and taking a long time to escape in the PM. For FS, when $\Pe$ is larger, since there are no such regions, the only effect is a more dominant advection than the diffusion, thus FS channel behavior is more sensitive to larger $\Pe$. {Importantly, as Pe increases (e.g., larger molecules with smaller diffusion coefficient), the peak value of the PDF curve for the FS channel increases, while that for the PM model decreases (as seen in Fig. \ref{fig:PDF})}, i.e., the PDF curve for the FS channel becomes narrower but the PDF curve for the PM becomes longer. This is because for the PM, the particles travel in all directions through the complex network of pores, thus generating a much larger “longitudinal dispersivity”, i.e, a higher equivalent diffusion in the longitudinal direction, proportional to $\Pe$\cite{Icardi:2014aa}. This means that, as $\Pe$ increases, the ISI of the PM channel increases but ISI of the FS channel decreases. Based on this, for the PM channel we expect the error performance and mutual information would become worse when $\Pe$ increases, which will be verified by the observations in Fig. \ref{fig:mutualError}.

%because the FS channel behavior is only affected by advection and diffusion, but the PM channel is also affected by dead-end pores or stagnant regions.
%the significant tailing that appears in the breakthrough curves (Fig.~\ref{fig:PDF}), when $\Pe$ is larger, changes the overall shape of the curve and not only its width. As already mentioned,

%Summarizing, we note that a larger P\'{e}clet number, leads to a longer PDF with a longer tail for the PM channel and, on the contrary, a narrower PDF with a shorter tail for the FS channel.%Thus, as $\Pe$ increases, the peak value difference of the PDF curves for the FS channel and the PM model becomes larger.
\vspace{-3mm}
\subsection{Performance Evaluation}
%In this subsection, we investigate the mutual information, the throughput, the error probability, and the diversity gain of the MC system via the PM and FS channel models.

\begin{figure}[!t]
\centering
\includegraphics[height=2in]{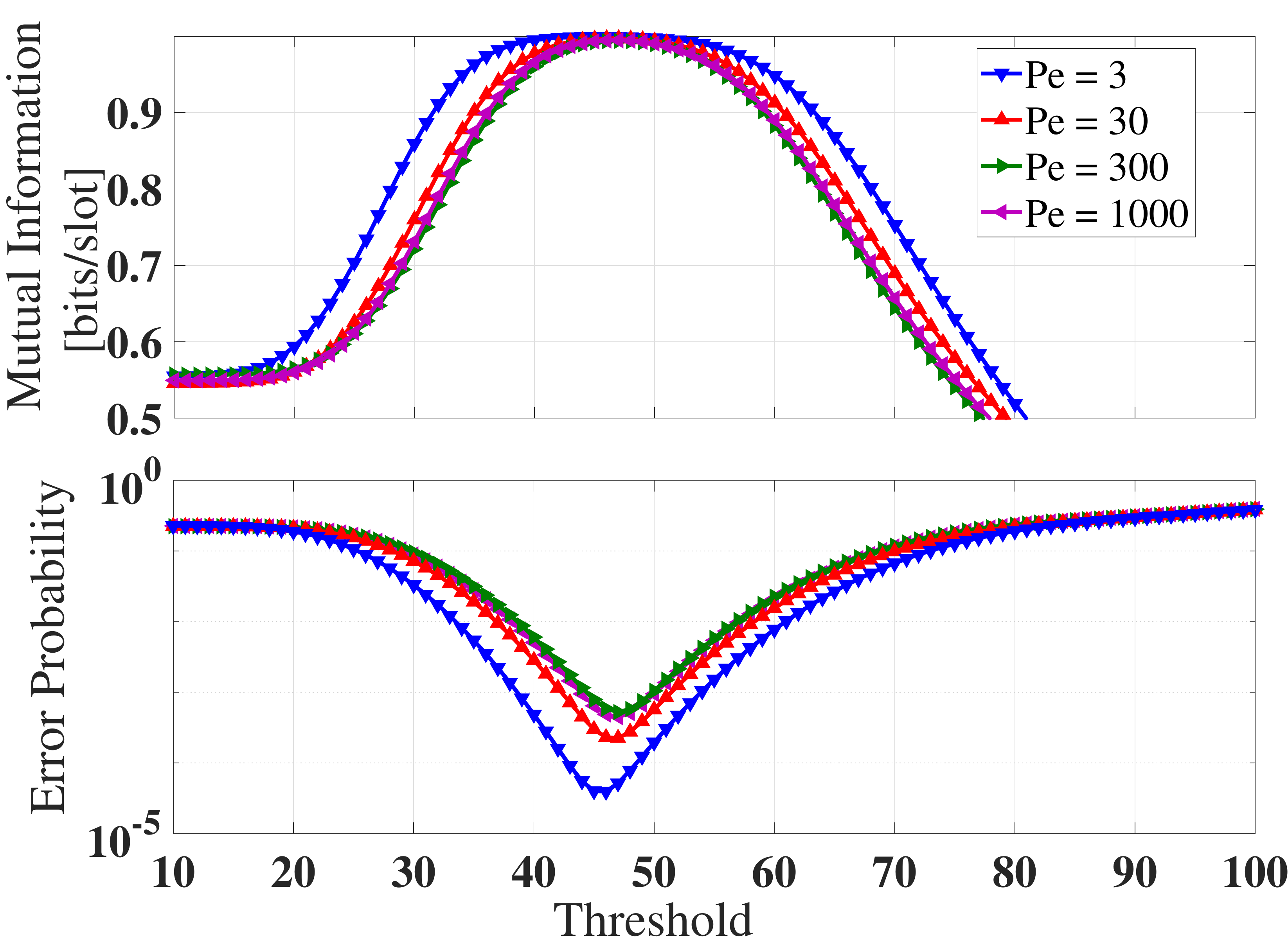}\vspace{-2mm}
\caption{{The average mutual information $I$ and the average error probability $Q$ of the MC system via the PM versus the threshold $\xi$ for different $\Pe$: $\Pe = 3,30,300,1000$. $N=100$ and $T=400\,\s$.}}
\label{fig:mutualError}
\vspace{-4mm}
\end{figure}
%\begin{figure}[!t]
%\centering
%\includegraphics[height=2.4in]{error_threshold_porous}
%\caption{The average error probability $Q$ of the MC system via the PM versus the threshold $\xi$ for different $\Pe$: $\Pe = 3,30,300,1000$. $N=100$, $T=400\,\s$, and $n=10$.}
%\label{fig:error}
%%\vspace{-4mm}
%\end{figure}

In Fig. \ref{fig:mutualError}, we show the average mutual information and the average error probability of the PM channel. Firstly, when $\xi=45$, $I$ is maximal (i.e., $I=1\,{\bit}/{\inter}$) and $Q$ is minimal, which numerically validates Corollary \ref{mutual}. Secondly, the average mutual information is smaller and the error probability is higher as $\Pe$ increases. This is because when $\Pe$ is higher, the tail of the channel response of the PM is longer, i.e., larger ISI, as we observed in Figs. \ref{fig:PDF} and \ref{fig:PDFCompare}.

%We observe that when $\xi=45$, Combining with Fig. \ref{fig:mutualError}(a), we find that the optimal threshold when the error probability is minimal is the same as that when the mutual information is maximal. We also observe that when $\Pe$ is larger, . This observation is expected since the tail of channel response of the PM is longer, i.e., larger ISI, as $\Pe$ increases, as illustrated in Figs. \ref{fig:PDF} and \ref{fig:PDFCompare}.

\begin{figure}[!t]
\centering
\includegraphics[height=2in]{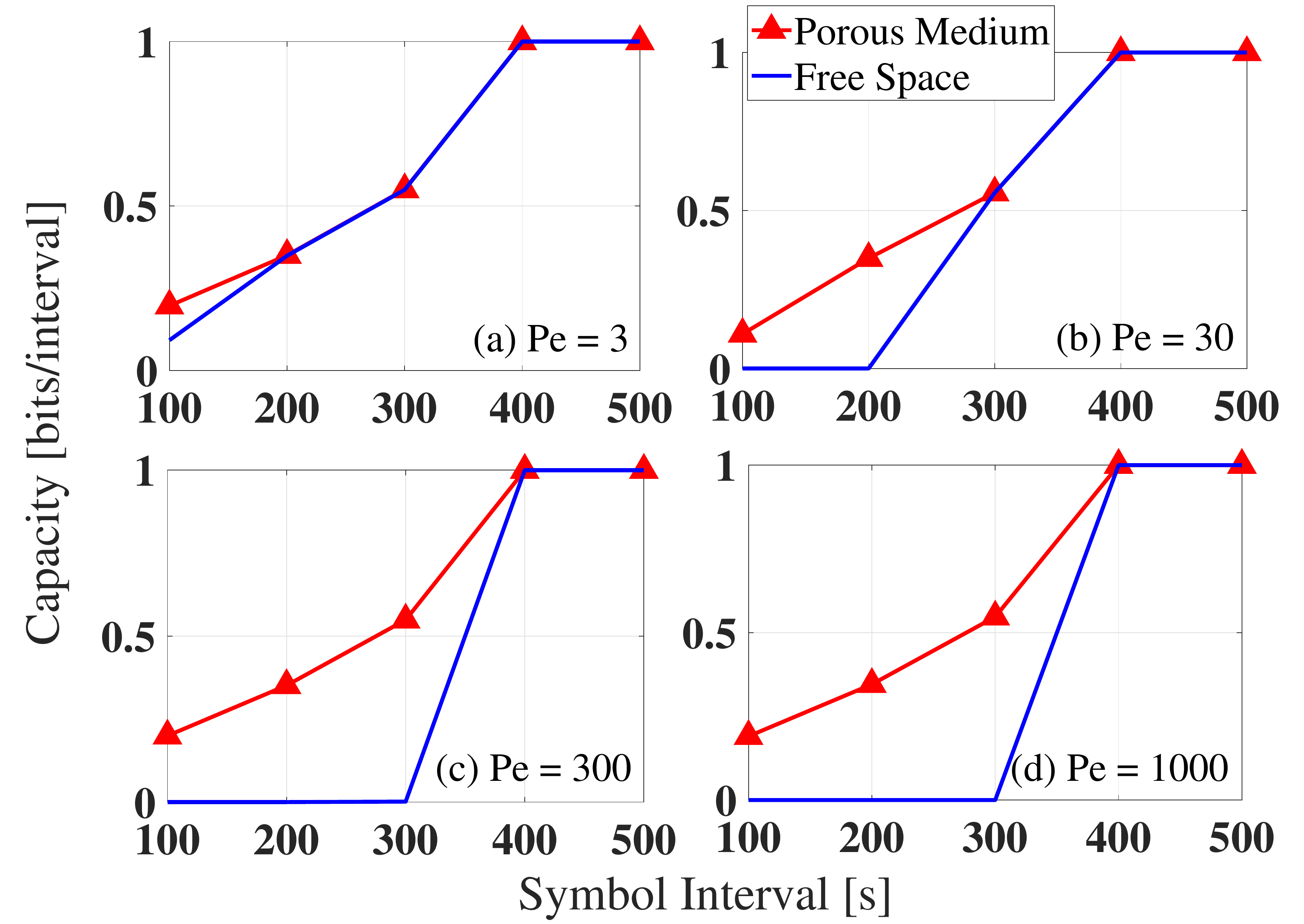}\vspace{-2mm}
\caption{The throughput $C$ of the MC system via the PM and FS channels versus the symbol slot $T$ with $N=10^5$ for different $\Pe$:  (a) $\Pe = 3$, (b) $\Pe = 30$, (c) $\Pe = 300$, and (d) $\Pe = 1000$.}
\label{fig:throughput}
\vspace{-4mm}
\end{figure}

In Fig. \ref{fig:throughput}, we show the throughput of the PM and FS channels. Firstly, for both channels and all $\Pe$, $C$ increases as $T$ increases and $ C=1\,{\bit}/{\inter}$ is achieved when $T\geq400\,\s$. This is because of a very small probability that a molecule arrives at $t\geq400\,\s$, as observed in Fig. \ref{fig:PDFCompare}. Secondly, the difference of $C$ between the PM and FS channels when $T\leq300\,\s$ becomes larger as $\Pe$ increases. This is because in Fig. \ref{fig:PDFCompare}, when $\Pe$ increases, the PM and FS channels diverge.

%the PDF curve for the FS channel becomes more narrow. This means that more molecules will arrive around $t=350\,\s$ and the throughput of FS channel for $T\leq300\,\s$ decreases.

\begin{figure}[!t]
\centering
\includegraphics[height=2in]{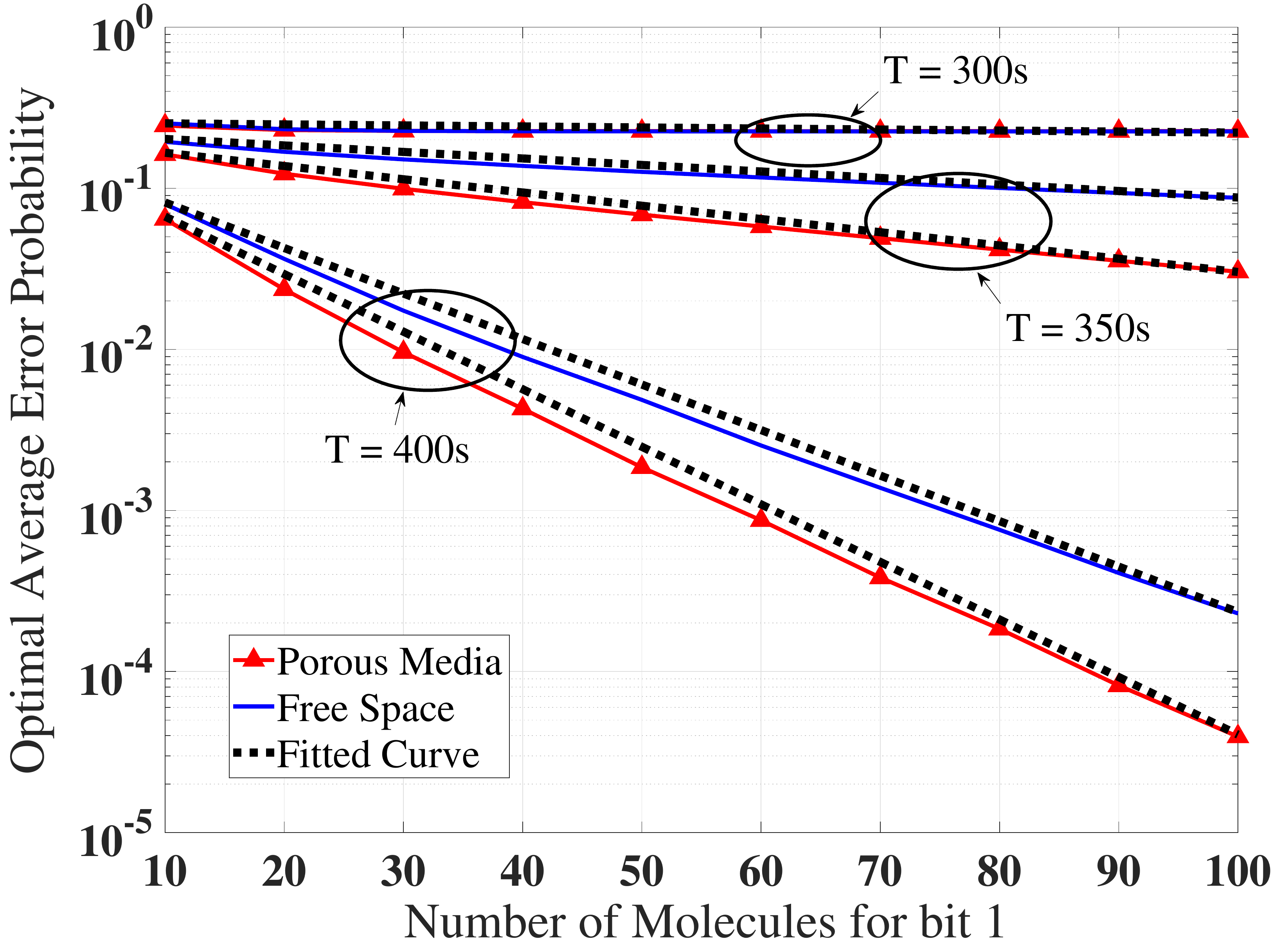}\vspace{-2mm}
\caption{{The optimal average error probability $Q^{\ast}$ of the MC system versus the number of molecules $N$ released for bit ``1'' for different symbol slots: $T=300\,\s$, $T=350\,\s$, and $T=400\,\s$ with $\Pe = 3$.}}
\label{fig:error_numMols}
\vspace{-4mm}
\end{figure}

\begin{table}[!t]
\renewcommand{\arraystretch}{1.1}
\centering
\caption{{Diversity Gain}}
\label{tab:table2}\vspace{-2mm}
\begin{tabular}{c|c||c|c|c|c}
\hline
\multicolumn{2}{c||}{Diversity Gain $\alpha$}                                                &  \bfseries $\Pe =3$ & \bfseries $\Pe = 30$ & \bfseries $\Pe = 300$ & \bfseries $\Pe = 1000$ \\ \hline \hline
\multirow{2}{*}{$T=300\,\s$}  & \begin{tabular}[c]{@{}c@{}}FS \end{tabular} &    0.0013    &  0.0055       &      0   &    0      \\ \cline{2-6}
                          & \begin{tabular}[c]{@{}c@{}}PM \end{tabular}   &   0.0009     &  0.0013       &  0.0010        &     0.0008\\ \hline
\multirow{2}{*}{$T=350\,\s$} & \begin{tabular}[c]{@{}c@{}}FS\end{tabular}    &  0.0089      &    0.0032     &   0.0048       &   0.0094        \\ \cline{2-6}
                          & \begin{tabular}[c]{@{}c@{}}PM \end{tabular} &  0.0186      &    0.0109     &     0.0144      &     0.0184      \\ \hline
\multirow{2}{*}{$T=400\,\s$} & \begin{tabular}[c]{@{}c@{}}FS\end{tabular}    &   0.0651     &   0.2689     &   0.8334      &    0.9487       \\ \cline{2-6}
                          & \begin{tabular}[c]{@{}c@{}}PM \end{tabular} &    0.0822    &  0.0651        &     0.0569     &  0.0585         \\ \hline
\end{tabular}
\vspace{-4mm}
\end{table}

In Fig. \ref{fig:error_numMols}, we plot the optimal average error probability versus the number of molecules released for bit ``1'' for different symbol slots. The considered symbol slots are around the detection time that maximizes the PM and FS channel responses based on Fig. \ref{fig:PDFCompare}. Firstly, $Q^{\ast}$ decreases when $N$ increases. %This means that, although both the ISI molecules and the intended signal molecules increase, the overall error decreases as the number of molecules increases.
We then see that error probability curves can be well approximated by the fitted curves, $Q^{\ast}\approx\exp(-\alpha N+\beta)$, where $\alpha$ and $\beta$ are obtained by solving $Q^{\ast}|_{N=10}=\exp(-\alpha 10+\beta)$ and $Q^{\ast}|_{N=100}=\exp(-\alpha 100+\beta)$. Thus, we can use the diversity gain $\alpha$ to quantify the decrease rate of $Q^{\ast}$ as $N$ increases. We present $\alpha$ for different $T$ and $\Pe$ in Table \ref{tab:table2}. %The diversity gain for FS and PM channels in (a) is $\alpha=0.065$ and $\alpha=0.082$, respectively, and (b) is $\alpha=0.009$ and $\alpha=0.019$, respectively. Using a similar method to that used for Fig. \ref{fig:error_numMols}, we obtain the diversity gain for different $\Pe$ with fixed symbol slot $T=350\,\s$ in Table \ref{tab:table1}.
We find that the PM achieves higher $\alpha$ than the FS channel for any $\Pe$ with $T=350\,\s$. This is because the decrease rate of $Q^{\ast}$ is affected by ISI. The PM has less ISI than the FS channel for these parameter values, based on the tails of the PDF curves of arrival time shown in Fig. \ref{fig:PDFCompare}.
%This may be explained by the fact that the transport of molecules in the FS channel is only affected by the diffusion and flow. The transport of molecules are also affected by the complex void and pore space in the PM, which leads to additional propagation paths and the higher diversity gain in the PM.

%Throughout this section, we find that PM channel behavior is much less sensitive to $\Pe$
%
%than the FS channel.
\vspace{-4mm}
\section{Conclusion}\label{sec:con}
%Since biological tissues can be modelled as porous media, molecular information delivery over porous media exists in biological environments.
We for the first time considered MC via a realistic PM channel, modeled as a 3D complex pore structure. %We numerically investigated the diversity gain which is referred to as the exponential decrease rate of the probability error as the number of released molecules increases.
Using fully resolved computational fluid dynamics results for the arrival time distribution, we explored the differences in channel characteristics between PM and FS channels and their impact on communication performance metrics (i.e.,
throughput, mutual information, error probability, and diversity gain) in both channels. Our results suggest that the reliability of a PM channel can be improved by decreasing $\Pe$, while opposite trends for a FS channel.

{Although the parameters (e.g., porosity, size, and topology) of different types of natural PM vary widely, their fundamental channel characteristics, i.e., the changing trends in the molecular arrival time distribution as Pe changes, are the same. This is because the key characteristic of molecular transport through the PM channel is that molecules may become trapped in the vicinity of solid grains, therefore taking some time to exit and causing non-trivial anomalous transport phenomena, such as long tails in the arrival time distributions. Our results reveal such changing trends in the molecular arrival time distribution and its impact on the different performance metrics of PM as Pe changes. These results provide useful guidelines for designing the optimal MC system through PM and predicting the system communication performance in a practical biological environment where Pe may change due to the instability of temperature and diffusion coefficients.}
%Notably, we find that as $\Pe$ increases, the tail of channel response of the PM is longer while that for the FS channel is shorter. This means opposite trends for the ISI which increases for the PM channel and decreases for the FS channel.
In our future work, we could analytically derive the arrival time distribution of a simplified, yet realistic, PM channel.

%A significant impact is also shown for the error performance and throughput of the channels. We also computed the channel throughput, mutual information, error probability, and diversity gain of the PM and FS channels.

%could include studying the adaptive threshold detection to improve the error performance of the system and

%transfer function
\vspace{-4mm}
%\bibliographystyle{IEEEtran}
%\bibliography{IEEEabrv,refs}


\clearpage

\begin{center}
\textsc{\normalsize{Supplementary Information}}\\
\end{center}

\appendices
\section{Derivation of Performance metrics}\label{app}

Due to the transport delay experienced by the molecules that arrive at the RX, the RX may receive the molecules released from the current and all previous symbol slots. Based on \eqref{breakthrough}, we obtain the probability that the molecule being released in the $k$th symbol slot arrives during the $n$th symbol slot, i.e., $F((n-k+1)T)-F((n-k)T)$. We denote $N^{\ob}_{n,k}$ as the number of molecules that arrive during the $n$th slot that were released at the beginning of the $k$th symbol slot. We then have $N^{\ob}_{n}=\sum_{k=1}^{n}N^{\ob}_{n,k}=\sum_{k=1}^{n-1}N^{\ob}_{n,k}+N^{\ob}_{n,n}$, where $\sum_{k=1}^{n-1}N^{\ob}_{n,k}$ is the ISI and $N^{\ob}_{n,n}$ is from the intended molecular signal. Since the molecules released in a given slot are transported independently and have the same probability to arrive during the $n$th slot, $N^{\ob}_{n,k}$ follows a binomial distribution, i.e.,
\begin{align}\label{binomial}
N^{\ob}_{n,k} \sim X_kB(N,F((n-k+1)T)-F((n-k)T)).
\end{align}

We note that modeling $N^{\ob}_{n,k}$ with the binomial distribution makes the analysis of $N^{\ob}_{n}$ cumbersome, since a sum of Binomial random variables (RVs) is not in general a Binomial RV. {Fortunately, $N^{\ob}_{n,k}$ can be accurately approximated by a Poisson distribution when $N$ is large and $F((n-k+1)T)-F((n-k)T)$ is small with $NF((n-k+1)T)-F((n-k)T)<10$}. By doing so, we rewrite $N^{\ob}_{n,k}$ as
\begin{align}\label{poisson1}
N^{\ob}_{n,k} \sim X_kP(N(F((n-k+1)T)-F((n-k)T))).
\end{align}

The sum of {independent} Poisson RVs is also a Poisson RV whose mean is the sum of the means of the individual Poisson RVs. As such, we have
\begin{align}\label{poisson}
N^{\ob}_{n} \sim P\left(\gamma\right).
\end{align}
where $\gamma=N\sum_{k=1}^{n}X_k(F((n-k+1)T)-F((n-k)T))$. In the following, we aim to derive $\Prob(N^{\ob}_{n}<\xi)$, since it lays the foundation for deriving all performance metrics in this paper. {Based on \eqref{poisson}, the CDF of the Poisson RV $N^{\ob}_{n}$ is written as}
{\begin{align}\label{Poisson,CDF}
\Prob(N^{\ob}_{n}<\xi|X_{1:n})= \sum_{j=1}^{\xi}\frac{\exp(-\gamma)\gamma^j}{j!}.
\end{align}}

We note that the the large number of summation terms in \eqref{Poisson,CDF} makes \eqref{Poisson,CDF} have very high computational complexity when $\xi$ is large. To facilitate the evaluation when $\xi$ is large, we further approximate $N^{\ob}_{n}$ as a Gaussian RV as follows:
\begin{align}\label{Guassian}
N^{\ob}_{n} \sim N(\gamma,\gamma),
\end{align}
where $\gamma=N\sum_{k=1}^{n}X_k(F((n-k+1)T)-F((n-k)T))$. {The Gaussian approximation for $N^{\ob}_{n}$ in \eqref{Guassian} is accurate when $\gamma>10$.} We define $X_{1:n} = \{X_1,X_2,\ldots,X_n\}$ as the subsequence of the symbols transmitted by the TX. Based on \eqref{Guassian}, we obtain the conditional CDF of the Gaussian RV $N^{\ob}_{n}$ for the given $X_{1:n}$ as
\begin{align}\label{Guassian,CDF}
\Prob(N^{\ob}_{n}<\xi|X_{1:n})=\frac{1}{2}\left(1+\erf\left(\frac{\xi-0.5-\gamma}{\sqrt{2\gamma}}\right)\right),
\end{align}
where $0.5$ is a continuity correction. Using \eqref{Poisson,CDF} or \eqref{Guassian,CDF}, we obtain the following conditional probabilities for the given $X_{1:n-1}$ as:
\begin{align}\label{prob00}
\Prob(Y_n=0|X_n=0,X_{1:n-1}) = \Prob(N^{\ob}_{n}<\xi|X_n=0,X_{1:n-1}),
\end{align}
\begin{align}\label{prob10}
&\Prob(Y_n=1|X_n=0,X_{1:n-1})\nonumber\\
= &\; 1-\Prob(N^{\ob}_{n}<\xi|X_n=0,X_{1:n-1}),
\end{align}
\begin{align}\label{prob01}
\Prob(Y_n=0|X_n=1,X_{1:n-1}) = \Prob(N^{\ob}_{n}<\xi|X_n=1,X_{1:n-1}),
\end{align}
and
\begin{align}\label{prob11}
&\Prob(Y_n=1|X_n=1,X_{1:n-1})\nonumber\\
= &\;1-\Prob(N^{\ob}_{n}<\xi|X_n=1,X_{1:n-1}).
\end{align}

{Using \eqref{prob00}-\eqref{prob11}, we first derive the conditional mutual information between channel input and output and the conditional symbol error probability given the subsequence of the previous symbols transmitted by the TX, $X_{1:n-1}$. To assess the overall system communication performance when transmitting different sequences of symbols, we then evaluate the average mutual information and the average symbol error probability over all realizations of $X_{1:n}$ and all symbol slots from 1 to $n$.}

\emph{\underline{Mutual Information}}: We derive the conditional mutual information between $X_n$ and $Y_n$ for the given $X_{1:n-1}$ as\footnote{We define $\Prob(\cdot|X_{1:n-1})\triangleq\Prob(\cdot)$, $I(\cdot|X_{1:n-1})=I(\cdot)$, and $H(\cdot|X_{1:n-1})\triangleq H(\cdot)$ in \eqref{mutualInfo}--\eqref{entropyHX1}.}
\begin{align}\label{mutualInfo}
&I(X_n;Y_n|X_{1:n-1})\nonumber\\
= &\;H(Y_n|X_{1:n-1})-H(Y_n|X_n,X_{1:n-1})~{{\bit}/{\inter}}.
\end{align}
where $H(\cdot)$ is the entropy. We derive $H(Y_n)$ as
\begin{align}\label{entropyH}
&\;H(Y_n|X_{1:n-1})\nonumber\\ =&-\Prob(Y_n=0|X_{1:n-1})\log_{2}\Prob(Y_n=0|X_{1:n-1})\nonumber\\
&-\Prob(Y_n=1|X_{1:n-1})\log_{2}\Prob(Y_n=1|X_{1:n-1}),
\end{align}
where $\Prob(Y_n=0|X_{1:n-1})$ and $\Prob(Y_n=1|X_{1:n-1})$ are written as
\begin{align}\label{Y0}
\Prob(Y_n=0|X_{1:n-1}) =&\; (1-P_1)\Prob(Y_n=0|X_n=0,X_{1:n-1})\nonumber\\
&+P_1\Prob(Y_n=0|X_n=1,X_{1:n-1})
\end{align}
and
\begin{align}\label{Y1}
\Prob(Y_n=1|X_{1:n-1}) =&\; (1-P_1)\Prob(Y_n=1|X_n=0,X_{1:n-1})\nonumber\\
&+P_1\Prob(Y_n=1|X_n=1,X_{1:n-1}),
\end{align}
respectively. We derive $H(Y_n|X_n,X_{1:n-1})$ as
\begin{align}\label{entropyHX}
H(Y_n|X_n,X_{1:n-1})=&\; (1-P_1)H(Y_n|X_n=0,X_{1:n-1})\nonumber\\
&+P_1H(Y_n|X_n=1,X_{1:n-1}),
\end{align}
where $H(Y_n|X_n=0,X_{1:n-1})$ and $H(Y_n|X_n=1,X_{1:n-1})$ are given by
\begin{align}\label{entropyHX0}
&H(Y_n|X_n=0,X_{1:n-1})\nonumber\\
=& -\Prob(Y_n=0|X_n=0,X_{1:n-1})\nonumber\\
&\times\log_{2}\Prob(Y_n=0|X_n=0,X_{1:n-1})\nonumber\\
&-\Prob(Y_n=1|X_n=0,X_{1:n-1})\nonumber\\
&\times\log_{2}\Prob(Y_n=1|X_n=0,X_{1:n-1}),
\end{align}
and
\begin{align}\label{entropyHX1}
&H(Y_n|X_n=1,X_{1:n-1})\nonumber\\
=& -\Prob(Y_n=0|X_n=1,X_{1:n-1})\nonumber\\
&\times\log_{2}\Prob(Y_n=0|X_n=1,X_{1:n-1})\nonumber\\
&-\Prob(Y_n=1|X_n=1,X_{1:n-1})\nonumber\\
&\times\log_{2}\Prob(Y_n=1|X_n=1,X_{1:n-1}),
\end{align}
respectively. We finally derive the average mutual information over all realizations of $X_{1:n-1}$ and all symbol slots from 1 to $n$ as
\begin{align}\label{AvermutualInfo}
I= \frac{1}{n}\sum_{k=1}^{n}\frac{\sum_{X_{1:k-1}\in\Psi_k} I(X_k;Y_k|X_{1:k-1})}{2^{k-1}}~{{\bit}/{\inter}},
\end{align}
where $\Psi_k$ is a set that includes all realizations of $X_{1:k-1}$.

\emph{\underline{Throughput}}: We derive the throughput, i.e., the maximal average mutual information, as
\begin{align}\label{throughput}
C = \underset{\xi}{\text{max}}~\frac{1}{n}\sum_{k=1}^{n}\frac{\sum_{X_{1:k-1}\in\Psi_k} I(X_k;Y_k|X_{1:k-1})}{2^{k-1}}~{{\bit}/{\inter}}.
\end{align}

\emph{\underline{Error Probability}}: We derive the symbol error probability in the $n$th slot for the given $X_{1:n-1}$ as
\begin{align}\label{error probability}
Q[n|X_{1:n-1}] =&\; (1-P_1)\Prob(Y_n=1|X_n=0,X_{1:n-1})\nonumber\\
&+P_1\Prob(Y_n=0|X_n=1,X_{1:n-1}).
\end{align}

We derive the average symbol error probability over all realizations of $X_{1:n-1}$ and all symbol slots from 1 to $n$ as
\begin{align}\label{AverError}
Q= \frac{1}{n}\sum_{k=1}^{n}\frac{\sum_{X_{1:k-1}\in\Psi_k} Q[k|X_{1:k-1}]}{2^{k-1}}.
\end{align}

%\emph{\underline{Diversity Gain}}: Since the transport of each molecule is independent and identically distributed, each molecule may experience different propagation paths. Thus, the channel can be seen as a multiple-input-multiple-output channel and the RX achieves diversity when $N$ molecules are released. We refer to the exponentially decreasing rate of the average error probability $Q$ as a function of increasing $N$ as the \emph{diversity gain}. That is to say, if we can well approximate $Q$ with a form of $Q\approx\exp(-\alpha N+\beta)$, then $\alpha$ is the diversity gain. We note that a similar definition of diversity gain was studied in \cite{Murin2018}, but our method for evaluating the diversity gain is different from \cite{Murin2018}. The diversity gain quantifies the decreasing rate of the error probability with $N$, i.e., a higher $\alpha$ means a higher decreasing rate and $\alpha=0$ means $Q$ does not change with $N$. Since an explicit expression for diversity gain is mathematically intractable, we use a data-fitting method to obtain $\alpha$ in Sec. \ref{Sec:results}.
%Since the analytical expression of $F(t)$ is not available, the explicit expression of the diversity gain $a$ is not available.

\section{Proof of Corollary \ref{limit}}\label{app1}
Since $Q$ is the sum of $Q[n|X_{1:n-1}]$ based on \eqref{AverError}, we need to prove that $Q^{\ast}[n|X_{1:n-1}]\to0$ when $N\to\infty$, where $Q^{\ast}[n|X_{1:n-1}]=\underset{\xi}{\text{min}}\;Q[n|X_{1:n-1}]$. Assuming $P_1 = \frac{1}{2}$, we first rewrite \eqref{error probability} as
\begin{align}\label{error probability1}
Q[n|X_{1:n-1}] =&\; \frac{1}{2}+\frac{1}{4}\left[\erf\left(\frac{\xi-0.5-(N(Y_1+Y_2))}{\sqrt{2(N(Y_1+Y_2))}}\right)\right.\nonumber\\
&\left.-\erf\left(\frac{\xi-0.5-NY_2}{\sqrt{2NY_2}}\right)\right],
\end{align}
where $Y_1=(F(T)-F(0))$ and $Y_2 = \sum_{k=1}^{n-1}X_k(F((n-k+1)T)-F((n-k)T))$. We then obtain the optimal $\xi$ that minimizes $Q[n|X_{1:n-1}]$. To this end, we take the first derivative of \eqref{error probability1} with respect to $\xi$ and solve the resultant equation to derive the optimal $\xi$ that minimizes $Q[n|X_{1:n-1}]$ as
\begin{align}\label{OptiThres}
\xi^{\ast}[n|X_{1:n-1}]= \frac{NY_1}{\ln\left({(Y_1+Y_2)}/{Y_2}\right)}.
\end{align}

Substituting \eqref{OptiThres} into \eqref{error probability1}, we write the optimal error probability $Q[n|X_{1:n-1}]$ as
\begin{align}\label{error probability2}
Q^{\ast}[n|X_{1:n-1}]=& \frac{1}{2}\!+\!\frac{1}{4}\!\left[\erf\!\left(\frac{\sqrt{N}\!A}{\sqrt{2(Y_1+Y_2)}}\right)\right.\nonumber\\
&\left.-\erf\left(\frac{\sqrt{N}\!B}{\sqrt{2Y_2}}\right)\right],
\end{align}
where
\begin{align}\label{con1}
A = \left(\frac{Y_1}{\ln\left({(Y_1+Y_2)}/{Y_2}\right)}\!-\!(Y_1+Y_2)\right)
\end{align}
and
\begin{align}\label{con2}
B = \left(\frac{Y_1}{\ln\left({(Y_1+Y_2)}/{Y_2}\right)}\!-\!Y_2\right).
\end{align}

If we can prove $A<0$ and $B>0$, then we have
\begin{align}\label{error probability3}
\lim_{N\to\infty} Q^{\ast}[n|X_{1:n-1}]=& \frac{1}{2}\!+\!\frac{1}{4}\!\left[\erf\!\left(-\infty\right)\right.\left.-\erf\left(\infty\right)\right]=0.
\end{align}

We now prove $A<0$ and $B>0$. Since $Y_1>0$ and $Y_2>0$, it is reasonably to assume $Y_1 = xY_2$, $x>0$. Using $Y_1 = xY_2$, we simplify the conditions $A<0$ and $B>0$ to $x/(1+x)-\ln(1+x)<0$ and $x-\ln(1+x)>0$, respectively. We find that $g(x) = x/(1+x)-\ln(1+x)$ is a decreasing function and $f(x) = x-\ln(1+x)$ is an increasing function with respect to $x$ since $g'(x)=-x/(1+x)^2<0$ and $f'(x)=1-1/(1+x)>0$ if $x>0$. By inspection, we also find $g(x)=0$ and $f(x)=0$ at $x=0$. Thus, we have $g(x)<0$ and $f(x)>0$ for $x>0$, which means $A<0$ and $B>0$. Thus, we verify that $Q^{\ast}[n|X_{1:n-1}]\to0$ when $N\to\infty$, which completes the proof.

\section{Proof of Corollary \ref{mutual}}\label{app2}
We first prove $I(X_n;Y_n|X_{1:n-1})\leq1\,{\bit}/{\inter}$. As per the Shannon entropy of probability distributions for single parties, we have $I(X_n;Y_n|X_{1:n-1})\leq \text{min}\{H(X_n|X_{1:n-1}),H(Y_n|X_{1:n-1})\}$. Based on definition of entropy, the maximal $H(X_n|X_{1:n-1})$ and $H(Y_n|X_{1:n-1})$ is $1\,{\bit}/{\inter}$ when $\Prob(X_1=0)=P_1=\frac{1}{2}$ and $\Prob(Y_1=0)=\frac{1}{2}$. Thus, the mutual information is bounded by $I(X_n;Y_n|X_{1:n-1})\leq1\,{\bit}/{\inter}$.

We then prove that $Q\to0$ is a sufficient condition for $I(X_n;Y_n|X_{1:n-1})=1\,{\bit}/{\inter}$. Based on \eqref{error probability}, $Q[n|X_{1:n-1}]\to0$ means $\Prob(Y_n=1|X_n=0,X_{1:n-1})\to0$ and $\Prob(Y_n=0|X_n=1,X_{1:n-1})\to0$. Applying these two expressions to \eqref{entropyH} and \eqref{entropyHX}, we obtain $I(X_n;Y_n|X_{1:n-1})=1\,{\bit}/{\inter}$, which proves $Q\to0$ is a sufficient condition. We finally prove that $Q\to0$ is a necessary condition for $I(X_n;Y_n|X_{1:n-1})=1\,{\bit}/{\inter}$. Since $H(Y_n|X_{1:n-1})\leq1$ and $H(Y_n|X_n,X_{1:n-1})\geq0$, thus $I(X_n;Y_n|X_{1:n-1})=1\,{\bit}/{\inter}$ is achieved only when $H(Y_n|X_{1:n-1})=1$ and $H(Y_n|X_n,X_{1:n-1})=0$. $H(Y_n|X_n,X_{1:n-1})=0$ means $H(Y_n|X_n=0,X_{1:n-1})=0$ and $H(Y_n|X_n=1,X_{1:n-1})=0$ based on \eqref{entropyHX}. There are four cases leading to $H(Y_n|X_n=0,X_{1:n-1})=0$ and $H(Y_n|X_n=1,X_{1:n-1})=0$ including:
\begin{enumerate}
\item $\Prob(Y_n=0|X_n=1,X_{1:n-1})=0$ and $\Prob(Y_n=1|X_n=0,X_{1:n-1})=0$;
\item $\Prob(Y_n=0|X_n=1,X_{1:n-1})=1$ and $\Prob(Y_n=1|X_n=0,X_{1:n-1})=0$;
\item $\Prob(Y_n=0|X_n=1,X_{1:n-1})=0$ and $\Prob(Y_n=1|X_n=0,X_{1:n-1})=1$;
\item $\Prob(Y_n=0|X_n=1,X_{1:n-1})=1$ and $\Prob(Y_n=1|X_n=0,X_{1:n-1})=1$.
\end{enumerate}
Since case 4) does not satisfy $\Prob(Y_n=0|X_n=1,X_{1:n-1})+\Prob(Y_n=1|X_n=0,X_{1:n-1})\leq1$, case 4) is not valid. Moreover, cases 2) and 3) result in $\Prob(Y_n=0|X_{1:n-1})=1$ and $\Prob(Y_n=1|X_{1:n-1})=1$, respectively, which leads to $H(Y_n|X_{1:n-1})=0$. Thus, they are not valid either. We note that only case 1) satisfies both $H(Y_n|X_{1:n-1})=1$ and $H(Y_n|X_n,X_{1:n-1})=0$ and case 1) leads to $Q\to0$. Thus, $Q\to0$ is a necessary condition. Therefore, we prove $Q\to0$ is a sufficient and necessary condition for $I(X_n;Y_n|X_{1:n-1})=1\,{\bit}/{\inter}$.

\end{document}